\documentclass[fleqn,usenatbib]{mnras}

\usepackage{newtxtext,newtxmath}
\usepackage[T1]{fontenc}
\usepackage{float}
\DeclareRobustCommand{\VAN}[3]{#2}
\let\VANthebibliography\thebibliography
\def\thebibliography{\DeclareRobustCommand{\VAN}[3]{##3}\VANthebibliography}

\usepackage{graphicx}	
\usepackage{amsmath}	
\newcommand{\source}{OJ 287}

\newcommand{\fermi}{{\it Fermi}-LAT}
\newcommand{\gray}{$\gamma$-ray}
\newcommand{\grays}{$\gamma$-rays}
\newcommand{\mmdc}{\textsf{MMDC}}

\usepackage[normalem]{ulem}

\title[Broadband emission of OJ 287]{Modeling of the time-resolved spectral energy distribution of blazar OJ 287 from 2008 to 2023: a comprehensive multi-epoch study}

\author[G. Harutyunyan et al.]{
G. Harutyunyan$^{1}$, \thanks{E-mail: gevorgharutyunyan97@gmail.com}
N. Sahakyan$^{1}$,
D. B\'egu\'e$^{2}$\\
$^{1}$ICRANet-Armenia, Marshall Baghramian Avenue 24a, Yerevan 0019, Armenia\\
$^{2}$Bar Ilan University, Ramat Gan, Israel
}

\date{Accepted XXX. Received YYY; in original form ZZZ}

\pubyear{\the\year{}}

\begin{document}
\label{firstpage}
\pagerange{\pageref{firstpage}--\pageref{lastpage}}
\maketitle

\begin{abstract}
We present a comprehensive analysis of the time-resolved spectral energy distributions (SEDs)
of the blazar OJ 287 over a 15-year period (2008–2023), using multi-wavelength data. In the $\gamma$-ray band, multiple flaring episodes were observed, with the strongest flare
reaching a peak flux of $(5.60\pm1.11)\times10^{-7}\:{\rm photons\:cm^{-2}\:s^{-1}}$ on MJD 55869.03
(04 November 2011). In the optical/UV band, the source was in an active state between MJD 57360 (04 December 2015) and
57960 (26 July 2017), during which the highest flux of $(1.07\pm0.02)\times10^{-10}\:{\rm erg\:cm^{-2}\:s^{-1}}$ was
observed on MJD 57681.23 (20 October 2016). In the X-ray band, both the flux and spectral index
exhibit variability.
To investigate the origin of the broadband emission from OJ 287, we systematically modeled 739
quasi-simultaneous SEDs using a leptonic model that self-consistently
accounts for particle injection and cooling. This analysis is possible thanks to the recent development of a surrogate neural-network-based model, trained on kinetic simulations. This innovative, time-resolved, neural network-based approach overcomes the limitations of traditional single-epoch SED modeling, enabling to explore the temporal evolution of key model parameters, such as the magnetic field strength, Doppler factor, and electron injection distribution, across different states of the source. We identified distinct emission states characterized by unique combinations of magnetic field \( B \), electron index \( p \), and Doppler boost \( \delta \), associated to different underlying mechanisms such as varying acceleration processes (e.g., shocks, turbulence) and magnetic confinement.
The analysis
provides insights into the jet physics processes, including particle acceleration
mechanisms and dynamic changes in the jet structure.
\end{abstract}




\section{Introduction}
Blazars are a type of active galactic nucleus (AGN) characterized by relativistic jets making a small angle (a few
degrees) with the line of sight of the observer \citep{1995PASP..107..803U}. As blazar jets are relativistic and
since the line of sight is making a small angle with the jet direction $\theta \sim 1/\Gamma$, with $\Gamma$
the Lorentz factor of the jet, the emission  is Doppler-boosted, making blazars among the most powerful extragalactic
sources, detectable even at high redshifts \citep[e.g.,][]{2017ApJ...837L...5A, 2020MNRAS.498.2594S,
2023MNRAS.521.1013S,2024MNRAS.528.5990S}. The emission from blazars, originating within the
jet, spans the entire electromagnetic spectrum, from radio waves to high-energy (HE; $\geq100$ MeV) and even very
high-energy (VHE; $\geq100$ GeV) \gray\ bands \citep{2017A&ARv..25....2P}. This broadband emission is characterized
by high-amplitude variability, which can sometimes occur on timescales as short as minutes \citep[e.g.,][]{2016ApJ...824L..20A,
2018ApJ...854L..26S, 2014Sci...346.1080A} or even seconds \citep[e.g.,][]{2007ApJ...664L..71A}. Recently, quasi-periodic
oscillations have been reported in the light curves
of some blazars \citep[e.g., ][]{2020ApJ...896..134P, 2023A&A...672A..86R}. Blazars are generally divided into
two categories based on their optical lines: BL Lacertae objects (BL Lacs), which have weak or absent optical
lines, and flat-spectrum radio quasars (FSRQs), which exhibit prominent broad optical lines \citep{1995PASP..107..803U}. 

The double-peaked spectral energy distribution (SED) of blazars is usually interpreted within the leptonic one-zone
or double-zone scenarios. In these scenarios, the electrons accelerated inside the jet produce the first peak
(typically in the optical/UV or X-ray bands) via synchrotron emission. The second peak is attributed to inverse
Compton scattering by the same electrons, occurring either in the same emitting region (one-zone) or in another
region (double-zone). The target photon field can originate from synchrotron radiation produced by the same
electrons \citep[synchrotron-self Compton model, SSC; ][]{1985A&A...146..204G, 1992ApJ...397L...5M,
1996ApJ...461..657B} or from photon fields external to the jet, such as photons from the accretion disk
\citep{1992A&A...256L..27D, 1994ApJS...90..945D}, photons reflected from the broad-line region
\citep{1994ApJ...421..153S}, or photons from the dusty torus \citep{2000ApJ...545..107B}. The synchrotron peak can also be used for another classification of blazars, defining low-synchrotron-peaked (LSP/LBL, $\nu_{p} <10^{14}$ Hz),
intermediate-synchrotron-peaked (ISP/IBL, $10^{14} < \nu_{p} <10^{15}$ Hz), and high-synchrotron-peaked
(HSP/HBL, $\nu_{p} > 10^{15}$ Hz) sources \citep{Padovani1995, Abdo_2010}.

Recently, with the advancement of multimessenger astrophysics, VHE neutrinos have been observed from the directions of the blazars TXS 0506+056 with the significance of the association estimated to be at the level of $3.0 \sigma$ \citep[$3.5\sigma$ for the neutrinos observed from the direction of TXS 0506+056 prior to the 2017 flaring episode][]{2018Sci...361..147I, 2018Sci...361.1378I, 2018MNRAS.480..192P} and PKS 0735+178 \citep{2021GCN.31191....1I, 2021ATel15112....1D, 2022ATel15290....1F}, which lies slightly outside the localization error region of IceCube-211208A but was undergoing a major multi-wavelength flare during the neutrino observations \citep{2023MNRAS.519.1396S}, making models involving hadronic interactions increasingly attractive.
Within hadronic models, the HE component is attributed either to the synchrotron emission of ultrahigh-energy protons
\citep{2001APh....15..121M} or to the secondaries generated through photo-pion and photo-pair interactions \citep{1993A&A...269...67M,
1989A&A...221..211M, 2001APh....15..121M, mucke2, 2013ApJ...768...54B, 2015MNRAS.447...36P, 2022MNRAS.509.2102G}.

The leptonic \citep[see e.g.,][]{2008ApJ...686..181F} and hadronic/lepto-hadronic \citep[see, e.g.,][]{2013ApJ...768...54B, 2018ApJ...863L..10A,2018ApJ...864...84K, 2018ApJ...865..124M, 2018MNRAS.480..192P, 2018ApJ...866..109S, 2019MNRAS.484.2067R,2019MNRAS.483L..12C, 2019A&A...622A.144S, 2019NatAs...3...88G, 2022MNRAS.509.2102G} models  have both been successfully applied to model either a single-epoch SED or a few, providing insights into the processes occurring within blazar jets.
While this is crucial for understanding the episodic activities in blazars, it is insufficient for a comprehensive
understanding, especially given the highly variable nature of their emission. The limitations in modeling
a large number of SEDs of the same object comes from (1) time-resolved data required to create simultaneous SEDs are not readily available and difficult to gather from scattered sources, (2)
the significant computational resources needed for such modeling. When the above theoretical models are applied in
a time-dependent manner—where particle injection and their evolution are followed to take into account particle cooling—it
becomes computationally expensive and practically difficult to integrate with traditional fitting tools.

In this context, we recently developed and provided to the community two major tools. First, the Markarian Multiwavelength Data Center\footnote{\url{www.mmdc.am}}
\citep[hereinafter \textsf{MMDC}][]{2024AJ....168..289S} is a state-of-the-art data-center which enables
the retrieval of time-resolved multiwavelength data from a large number of blazars. Through \mmdc{}, the
construction of simultaneous SEDs for the same blazar over different periods, facilitates the study of dynamic
changes in the radiative components over time. Second, in \cite{2024ApJ...963...71B} and \cite{2024ApJ...971...70S},
a novel method was introduced for emulating blazar SEDs with a neural networks, and using
these SED models to fit the data. This approach bypasses the time-consuming model calculation, which
enables fits to be performed. Indeed, with this enhancement, Bayesian fitting tools
can be used to retrieve the free model parameters that statistically better describe the data. The availability of such data,
combined with advanced fitting methods, significantly enhances our ability to understand the complex physics of blazars, propelling blazar analysis into a new, time-resolved, data-rich era.

Taking advantage of the availability of both extensive data and advanced modeling tools provided by \textsf{mmdc} (both publicly available),
we have initiated a project to study the origin of multiwavelength emission from blazars through comprehensive
modeling of their SEDs over different epochs. The results of this project, titled Modeling of
time-resolved Spectral Energy Distributions of blazars (MSED), will
be made available online shortly. This approach enables us to constrain the model parameters during all time
periods in which data are available and investigate their evolution over time. Consequently, the observed variability
in flux can be attributed to changes in the physical conditions of these objects, which are retrieved in our modeling
providing a deeper understanding of the processes occurring within blazar jets and their temporal evolution. The first blazar we considered is OJ 287, a LBL at a redshift of \( z = 0.306 \) \citep{1985PASP...97.1158S}.
In the optical band, observations of OJ 287 reveal a quasi-periodic behavior with an approximately 12-year cycle, which has been
explained either by the presence of a binary black hole system \citep[e.g.,][]{1988ApJ...325..628S, 1997ApJ...484..180S,
2012MmSAI..83..219V} or through helical jet models \citep[e.g.,][]{2013A&A...557A..28V}. Due to this behavior OJ 287 was
always under extensive multiwavelength campaign resulting in a large amount of data well sampled in time. For this reason,
OJ 287 was selected and the results of our analysis are presented in this paper.

This paper is organized as follows. Section \ref{analy} presents the broadband data analysis conducted for
this study.
The "time-resolved" SEDs obtained over different epochs are described in Section \ref{SED}, while their modeling
is detailed in Section \ref{orig}. The results are presented in Section \ref{res}, and the conclusions along with
discussion are provided in Section \ref{disc}.

\section{Data analysis}
\label{analy}
The time resolved SEDs are build using the data from the optical/UV, X-ray, and \gray\ bands retrieved from \mmdc{} \citep{2024AJ....168..289S},
a novel platform where time-resolved data from blazar observations are analyzed and made available to the scientific community. The methodology
for analyzing data across different bands is detailed in this section. 

\subsection{Optical/UV data}
In the optical/UV band, data from the Ultra-Violet and Optical Telescope \citep[UVOT;][]{2005SSRv..120...95R} on board
the Neil Gehrels Swift Observatory \citep[][hereafter Swift]{2004ApJ...611.1005G}, were analyzed, alongside
data retrieved from the All Sky Automated Survey for SuperNovae \citep[ASAS-SN;][]{2017PASP..129j4502K}, the Zwicky
Transient Facility \citep[ZTF;][]{2019PASP..131a8002B}, and the Panoramic Survey Telescope and Rapid Response System
\citep[Pan-STARRS;][]{2002SPIE.4836..154K}, to investigate the emission of \source\ in these bands.

Swift-UVOT observed \source\ in the optical (V: 500-600 nm, B: 380-500 nm, and U: 300-400 nm) and UV (W1: 220-400 nm,
M2: 200-280 nm, and W2: 180–260 nm) filters. The UVOT data from all blazars observed by Swift are analyzed and made
publicly available through \mmdc{} \citep{2024AJ....168..289S}. Among the data from \source\ observations, 890 observations
fall within the period considered in the current study (from August 2008 to July 2023). The detailed analysis of the Swift
UVOT data is presented in \citet{2024AJ....168..289S} and is briefly summarized here. The observations
were reduced using the standard procedure with HEAsoft version 6.29 and the latest release of the HEASARC CALDB. Source
counts were extracted from a 5 arcsec circular region, while background counts were taken from a 20 arcsec region away
from the source. The photon count-to-flux conversion was based on UVOT calibration \citep{2008MNRAS.383..627P}, with
corrections for extinction applied using the reddening coefficient ${E(B-V)=0.0282}$ from the Infrared Science Archive\footnote{http://irsa.ipac.caltech.edu/applications/DUST/}.
The optical and UV light curves are shown in Fig. \ref{fig:MWLC} panels (d) and (e).

In addition to the Swift UVOT data, archival data in various optical filters were also used.
Specifically, V and g band data from the ASAS-SN \citep{2017PASP..129j4502K} public archive were retrieved by performing
a search within a 5-arcsecond radius around the source position. Similarly, data in the g, r, and i bands, as well as in
the g, r, i, z, and y filters, were retrieved by conducting a cone search within a 5-arcsecond radius in the public archives
of ZTF \citep{2019PASP..131a8002B} and Pan-STARRS \citep{2002SPIE.4836..154K}. After the search, additional verification is performed to check whether nearby sources are present, as they could potentially cause confusion.   These data were corrected for extinction and
are also available through \mmdc{} \citep{2024AJ....168..289S}. The light curves from
ASAS-SN, ZTF, and Pan-STARRS observations of \source\ are presented in Fig. \ref{fig:MWLC}, panel (f).

The optical/UV light curves shown in panels (d), (e), and (f) of Fig. \ref{fig:MWLC} demonstrate that the source occasionally flares in these bands. In its brightest state, the flux increases by a factor of 4-5 compared to the average state, reaching up to highest flux of $(1.07\pm0.02)\times10^{-10}\:{\rm erg\:cm^{-2}\:s^{-1}}$ observed on MJD 57681.23 in the M2 filter. Notably, significant flaring activity in the optical/UV bands was observed between MJD 57360 and 57960, following the \gray\ brightening around MJD $\sim57100$ (see below).

In addition, panel (g) of Fig. \ref{fig:MWLC} presents infrared (IR) data from \source\ observations conducted with the Wide-field Infrared Survey Explorer \citep[WISE;][]{2010AJ....140.1868W} and its reactivated mission, the Near-Earth Object Wide-field Infrared Survey Explorer \citep[NEOWISE;][]{2011ApJ...731...53M}. WISE/NEOWISE data were retrieved from \mmdc{} \citep{2024AJ....168..289S}, considering only measurements in the 3.4 and 4.6$\mu$m bands. WISE observations provide also a single-epoch measurement at 12 and 22$\mu$m, which is not informative enough to include in the light curve. Panel (g) of Fig. \ref{fig:MWLC} also shows that the source exhibited periods of high emission in the IR band. The highest flux in the 4.6$\mu$m band was $(6.44\pm0.07)\times10^{-11}\,{\rm erg\,cm^{-2}\,s^{-1}}$ observed on MJD 57695.2, while for the 3.4$\mu$m band it was $(6.42\pm0.09)\times10^{-11}\,{\rm erg\,cm^{-2}\,s^{-1}}$ observed on MJD 57695.4. These IR observations are specially important as \citet{2011ApJ...740L..48M} demonstrated that in the 3.4-4.6-12 $\mu$m color-color diagram blazars cover a distinct region while \citet{2016ApJ...827...67M} showed correlation between the mid-IR colors and the \gray\ index. Moreover, \citet{2024ApJ...963...48G} showed that 3.4 and 4.6$\mu$m spectral slope between 3.4 and 4.6$\mu$m can predict the peak frequency and maximum intensity of the synchrotron component of blazar SEDs.

\subsection{X-ray data}
\source\ was monitored in the X-ray band with Swift XRT in the 0.3-10 keV band and with NuSTAR between 3 and 79 keV. The data from both Swift XRT and NuSTAR were reduced to study the X-ray emission properties of \source.
\subsubsection{Swift XRT}
Swift XRT observed \source\ 890 times up until July 4, 2023 (the average exposure time of the Swift XRT observations being 1.13 ks). These data, along with all other data from blazar observations,
are analyzed and made accessible through \mmdc{}. The details of the X-ray data analysis are provided in
\citet{2024AJ....168..289S} and are briefly summarized here. The analysis was performed using the
{\it Swift\_xrtproc} tool, which is based on the XRT Data Analysis Software (XRTDAS) and the spectral and imaging analysis
tools XSPEC and XIMAGE, and follows standard procedures \citep{2021MNRAS.507.5690G}. For each observation, this tool
downloads the archival data, performs standard data cleaning and filtering procedures, and produces calibrated data
products. The X-ray spectrum for each observation is then extracted by selecting events within a circular region of
20-pixel radius ($\sim47$ arcsec), with background counts extracted from a nearby source-free circular region of 40-pixel radius. In addition, a verification of the source count rate was performed, and when the count rate exceeds \(0.5\:{\rm count\:s^{-1}}\), a pile-up correction was applied \citep[for details, see][]{2021MNRAS.507.5690G}. The data are grouped using the GRPPHA tool to include at least one count in each energy bin, which are then
loaded into XSPEC \citep{1996ASPC..101...17A} for spectral fitting using Cash statistics \citep{1979ApJ...228..939C}.
Both power-law and log-parabola models (using fixed Galactic value of $2.38\times10^{20}\:{\rm cm^{-2}}$) were assumed for the spectrum of the source, estimating the X-ray photon index
and flux in various bands. During the fitting, no additional absorption in the rest frame of the sources was taken into account.

The 2-10 keV X-ray flux variation is shown in panel (b) of Fig. \ref{fig:MWLC}, and the time-variation of the photon index is depicted in panel (c). The mean flux
in the X-ray band is approximately $\sim4.65\times10^{-12}\:{\rm erg\:cm^{-2}\:s^{-1}}$, although the flux occasionally
exceeds $10^{-11}\:{\rm erg\:cm^{-2}\:s^{-1}}$. The maximum flux of $(1.52\pm0.21)\times10^{-11}\:{\rm erg\:cm^{-2}\:s^{-1}}$
was observed on MJD 54875.91. The photon index also varies over time, with an average value of $\sim1.99$. The hardest
index of $1.22\pm0.22$ was observed on MJD 54780.28, during which the 2-10 keV flux was as high as
$(1.33\pm0.04)\times10^{-11}\:{\rm erg\:cm^{-2}\:s^{-1}}$. There are also periods when the source is in a bright state,
with the 2-10 keV flux exceeding $10^{-11}\:{\rm erg\:cm^{-2}\:s^{-1}}$, while the photon index remains relatively soft
($\geq2.5$). For instance, on MJD 57786.36, the 2-10 keV flux was $(1.28\pm0.07)\times10^{-11}\:{\rm erg\:cm^{-2}\:s^{-1}}$
with a photon index of $2.50\pm0.06$. The change in the photon index can be attributed to the dominant component contributing to the X-ray band. When the synchrotron component extends into the X-ray band, it results in a softer spectrum, whereas if the X-ray band corresponds to the inverse Compton component, it produces a harder spectrum.

\subsubsection{NuSTAR}
Operating in the 3-79 keV energy range with two focal plane modules (FPMs), FPMA and FPMB, NuSTAR has observed
\source\ four times: on 9 April 2017, 4 May 2020, 22 March 2022 and 1 December 2022.
These observations amount to a total of 126.96 ks of observation time. Similar to Swift XRT data, all blazar data
observed by NuSTAR are available through \mmdc{} \citep{2024AJ....168..289S}. A brief description of the analysis is
as follows. The raw data are accessed, downloaded, and analyzed using the \textit{NuSTAR\_Spectra} tool
\citep{2022MNRAS.514.3179M}. This tool downloads the data and applies standard analysis techniques to generate
high-level calibrated scientific products using the \textit{nuproducts} routine, which are then ready for spectral
analysis. The source counts are extracted from a circular region, while the background counts are extracted from an
annulus centered on the source. The radii of these regions are automatically optimized based on the source count
rate (with a higher count rate resulting in a larger radius and a lower count rate resulting in a smaller radius).
Subsequently, assuming both power-law and log-parabola spectra (taking into account also the Galactic absorption), the best-fit parameters are obtained using the XSPEC
package \citep{1996ASPC..101...17A} with the Cash statistics \citep{1979ApJ...228..939C}.

The analysis reveals slight variability in the X-ray emission from \source\ in the hard X-ray band across different
observations. Specifically, the 3-10 keV flux during the 2017 and 2020 observations was
$(3.64\pm0.06)\times10^{-12}\:{\rm erg\:cm^{-2}\:s^{-1}}$ and $(3.51\pm0.08)\times10^{-12}\:{\rm erg\:cm^{-2}\:s^{-1}}$,
respectively, while in March and December 2022, it was $(2.15\pm0.11)\times10^{-12}\:{\rm erg\:cm^{-2}\:s^{-1}}$ and
$(4.83\pm0.08)\times10^{-12}\:{\rm erg\:cm^{-2}\:s^{-1}}$, respectively. The photon index remained relatively consistent
during the 2022 observations ($\Gamma=1.90\pm0.11$ and $\Gamma=1.82\pm0.04$) and the 2017 observation ($\Gamma=2.06\pm0.04$).
However, X-ray emission with a significantly softer index was observed during the 2020 observation,
with $\Gamma=2.42\pm0.06$. This softer index coincides with Swift XRT observation 00035905053, which also reported a
similarly soft index of $2.49\pm0.14$. 
\subsection{$\gamma$-ray data}
The \gray\ data from Fermi Large Area Telescope (\fermi) are used to investigate
the emission from \source. The LAT is a pair-conversion
telescope sensitive to \grays\ in the energy range from 20 MeV to 300 GeV. By default, it performs an all-sky survey
every 3 hours. For more details on the \fermi\ instrument, see \citet{2009ApJ...697.1071A}.

The data analysis performed in this work adheres to the standard point-like source analysis procedures recommended
by the \fermi\ collaboration. The complete analysis methods and strategy are detailed in \citet{2024AJ....168..289S}
and are briefly summarized as follows. The analysis was performed with \textit{fermitools} version 2.0.8, using the
P8R3\_SOURCE\_V3 instrument response functions. Events within the energy range of 100 MeV to 300 GeV were selected
within a $12^\circ$ radius centered on the \gray\ location of \source\ (RA$=133.70$, Dec$=20.10$). The events observed
from August 4, 2008,and July 4, 2023 (MET 239667417-710178221), were considered, retrieving only those classified as
{\it evclass=128} and {\it evtype=3}. All events with zenith angles greater than $90^\circ$ were excluded to avoid \grays\
produced by Earth-limb effects. During the likelihood analysis, sources within a $17^\circ$ radius centered on the target,
as listed in the \fermi\ fourth source catalog (4FGL) incremental version \cite[DR 3;][]{2022ApJS..260...53A}, were
considered. The spectral parameters of sources within a $12^\circ$ radius were allowed to vary freely, while those
outside this region were fixed to the values reported in the 4FGL catalog. The model also included Galactic and
extragalactic background models, utilizing the latest available files, gll\_iem\_v07 and iso\_P8R3\_SOURCE\_V3\_v1,
respectively. The model parameters were optimized using a maximum likelihood analysis, with the test statistic (TS)
defined as $TS = {\rm 2(lnL_1-lnL_0)}$, where $L_1$ and $L_0$ represent the maximum likelihoods with and without the
source, respectively \citep{1996ApJ...461..396M}, to quantify the significance of the \gray\ emission.

To investigate the variability of \source\ in the \gray\ band, a light curve was generated using the adaptive
binning method \citep{2012A&A...544A...6L}. This method defines time intervals in the light curve based on the
uncertainty in flux estimation rather than using fixed intervals. The advantage of flexible time bins is that
they allow for narrow time intervals when the source is in a bright emission state, while extending over longer
periods when the source is in a quiescent state. This enables the detection of short-scale variations in the
\gray\ flux of \source, providing a comprehensive view of its temporal behavior. This method has been successfully
applied to study \gray\ emission from other blazars \citep[see, e.g.,][]{2013A&A...557A..71R, 2016ApJ...830..162B, 2017MNRAS.470.2861S, 2017A&A...608A..37Z, 2017ApJ...848..111B, 2018ApJ...863..114G, 2018A&A...614A...6S, 2021MNRAS.504.5074S, 2022MNRAS.517.2757S, 2022MNRAS.513.4645S}. To compute the adaptively binned light curve of \source, the entire
observation period was initially divided into smaller intervals that satisfy the condition that the flux estimation
uncertainty above the optimal energy of $>E_{\rm opt}=237.6$ MeV is 20\%. For all
identified periods, a standard unbinned likelihood analysis was then applied using the same criteria as described
earlier, but modeling the overall spectrum of \source\ as a power law, as for shorter periods it provides a better
explanation of the \gray\ spectrum.

\begin{figure*}
    \centering
    \includegraphics[width=1\linewidth]{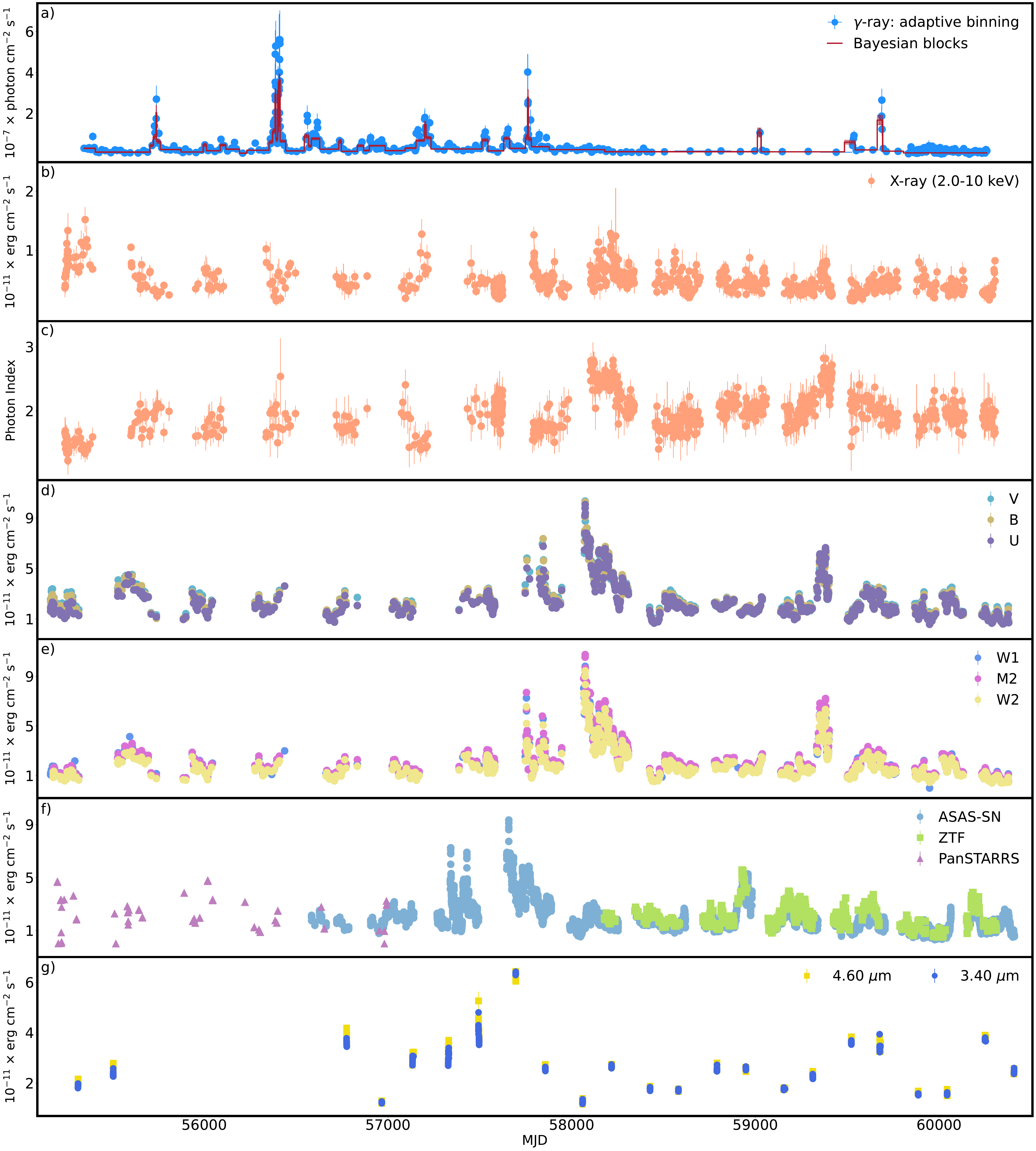}
    \caption{
    The multiwavelength light curve of \source\ from August 4, 2008, to September 9, 2023. Panel a) shows the
    adaptively binned \gray\ light curve (blue) together with the Bayesian intervals (red). Panels b) and c)
    show the X-ray flux in the 2.0–10 keV range and the photon index in the 0.3–10 keV range, respectively.
    Swift UVOT fluxes in the optical band (V, B, U filters) are shown in panel d), while the UV band (W1, M2,
    and W2 filters) is presented in panel e). The optical data from SAS-SN, ZTF, and PanSTARRS are shown in panel f). Panel g) shows the IR flux in the 3.4 and 4.6$\mu$m bands from WISE/NEOWISE observations.}
    \label{fig:MWLC}
\end{figure*}

The \gray\ light curve of \source\ is presented in panel (a) of Fig. \ref{fig:MWLC}. In this band, the source underwent several
flaring episodes, with the most intense flare occurring between MJD 55827 and 55874 (09 September 2011 to 09 November 2011),
during which the highest flux of $(5.60\pm1.11)\times10^{-7}\:{\rm photon\:cm^{-2}\:s^{-1}}$ was observed on MJD 55869.03
(04 November 2011). Other flaring periods were also detected, albeit with lower magnitudes. During this major
flare, the photon index slightly hardened, generally being above 2.0 and occasionally even
below 1.90. The hardest index of $1.55\pm0.19$ was observed on MJD 55867.14 (02 November 2011).

\section{Broadband SEDs of \source\ and its evolution in time}\label{SED}

The rich dataset described in the previous section provides a wealth of information
about the emission components of \source\ across different bands and their time evolution \citep[for other multi-wavelength
variability studies of \source, see also][]{2024MNRAS.532.3285Z, 2024ApJ...960...11K, 2024ApJS..270...22W, 2025ApJ...979..210Z}.
While a single SED snapshot or a time-average SED could offer valuable insights into the
underlying physical processes, a comprehensive understanding of the emission mechanisms requires the analysis of all
available time-resolved SEDs, ideally obtained from simultaneous multi-wavelength observations. This approach
allows for capturing the changes that lead to variability
and studying the evolution of the emission over time. Therefore, analyzing as many simultaneous SEDs as possible
is crucial for achieving a more complete and accurate characterization of blazar behavior.

To obtain a comprehensive picture of the evolution of emission from \source, an SED/LC animation was generated.
This method has been successfully applied to investigate time-resolved emissions from BL Lac \citep{2022MNRAS.513.4645S},
3C 454.3 \citep{2021MNRAS.504.5074S}, and CTA 102 \citep{2022MNRAS.517.2757S}. To create this animation, in
addition to the data presented in the previous section, archival multiwavelength data were retrieved from \mmdc{}
\citep{2024AJ....168..289S}, which aggregates data from blazar observations in the radio-to-X-ray bands across numerous
catalogs, as well as IR data from the Wide-field Infrared Survey Explorer (WISE) and NEOWISE surveys
\citep{2014ApJ...792...30M}. The adaptively binned light curve was then divided into segments (indicated by the red
line in panel (a) of Fig. \ref{fig:MWLC}), corresponding to Bayesian Blocks that exhibit
similar flux levels. For each period (i.e., each Bayesian Block), a detailed spectral analysis was conducted
to obtain the spectrum in the HE \gray\  band observed by Fermi-LAT. These periods are shorter when the
source is in an active emission state and longer when it is in a quiescent emission state. The analysis of these Bayesian
Blocks shows that the minimal flux level, selected from the adaptively binned light curve, corresponds to
$(1.37\pm0.65)\times10^{-8}\:{\rm photon\:cm^{-2}\:s^{-1}}$, observed during the period MJD 55626.54 – 55670.87, while the highest flux of $(8.08\pm0.57)\times10^{-7}\:{\rm photon\:cm^{-2}\:s^{-1}}$ was recorded between MJD 55863.79 – 55870.82. Similarly, the softest photon index of $-2.92 \pm 0.21$ was observed between MJD 55408.29 – 55427.64, while the hardest index of $-1.79 \pm 0.21$ was observed between MJD 55626.54 – 55670.87. This demonstrates that the Bayesian Blocks method effectively captures different states of the source \citep[see also][for the \gray\ spectrum of \source\ based on the analysis of long-term accumulated data from \fermi\ ]{2021A&A...654A..38P, 2024OJAp....7E..64P}.

All analyzed and retrieved archival data with known observation
times were subsequently binned into Bayesian Blocks and displayed alongside the HE \gray\ data.
Within each Bayesian Block, an adaptive time scan was performed to identify as many simultaneous data points as
possible. By sequentially displaying the resultant SEDs, the animation effectively illustrates the temporal evolution of the observed emission of \source. For more details on generating SED/LC animations, see \citet{2024AJ....168..289S}.

The SED/LC animation generated for \source\ is available on \href{https://youtu.be/BFNP3sxymT4} {\nolinkurl{youtube}}
and can also be accessed via \mmdc{}  under "SED/LC Animation". This animation illustrates the significant variability
in emission across different bands, along with spectral changes that shift the peaks of the emission components. For
instance, the flux increases by factors of 23, 46, and 70 at frequencies of $10^{15}$ Hz, $10^{18}$ Hz,
and $10^{24}$ Hz, respectively. The animation also demonstrates the evolution of different components over time; for
example, around MJD 57790, the X-ray component softened, shifting the synchrotron component into the X-ray band.
Additionally, it shows that the HE component occasionally hardens, shifting its peak to higher frequencies.

\begin{figure*}
    \centering
    \includegraphics[width=0.48\linewidth]{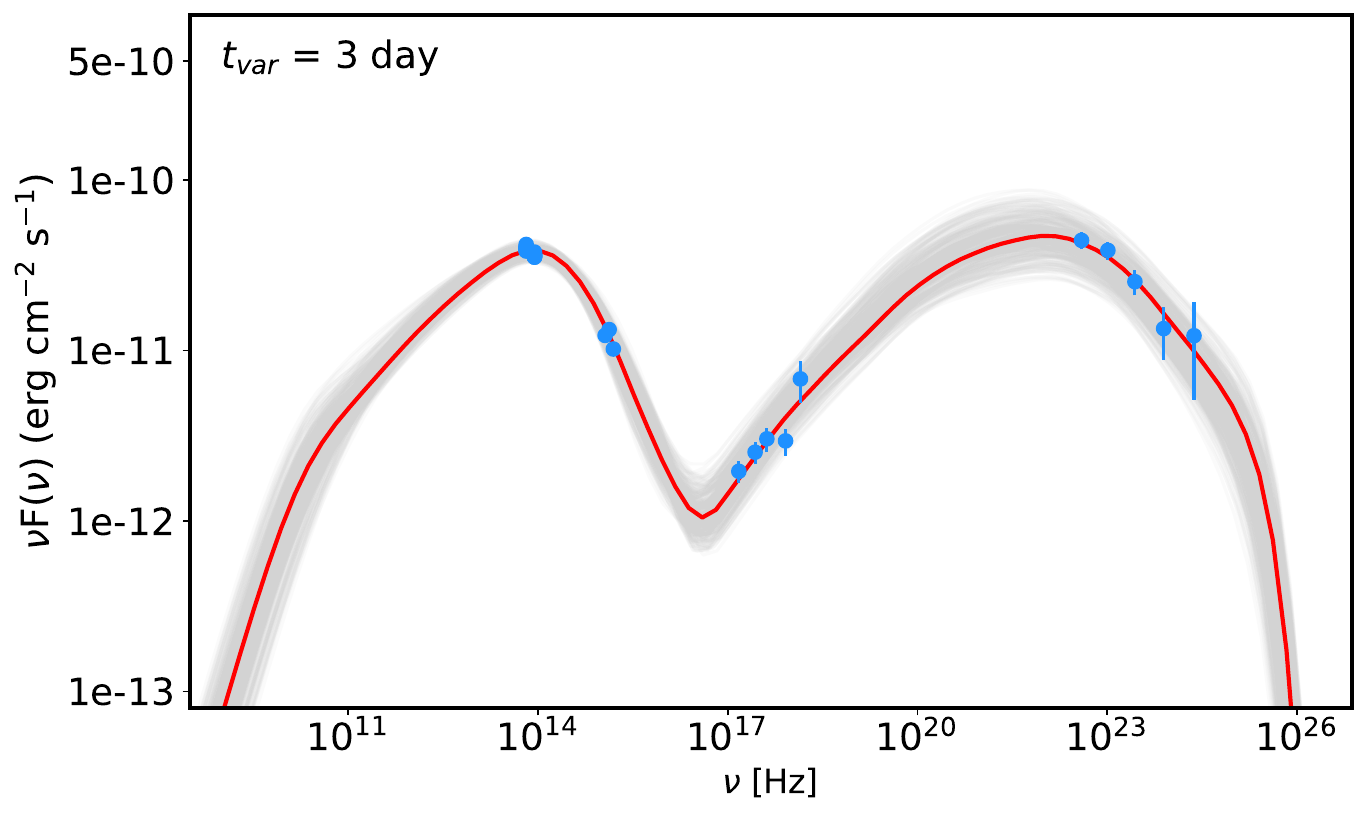}
    \includegraphics[width=0.48\linewidth]{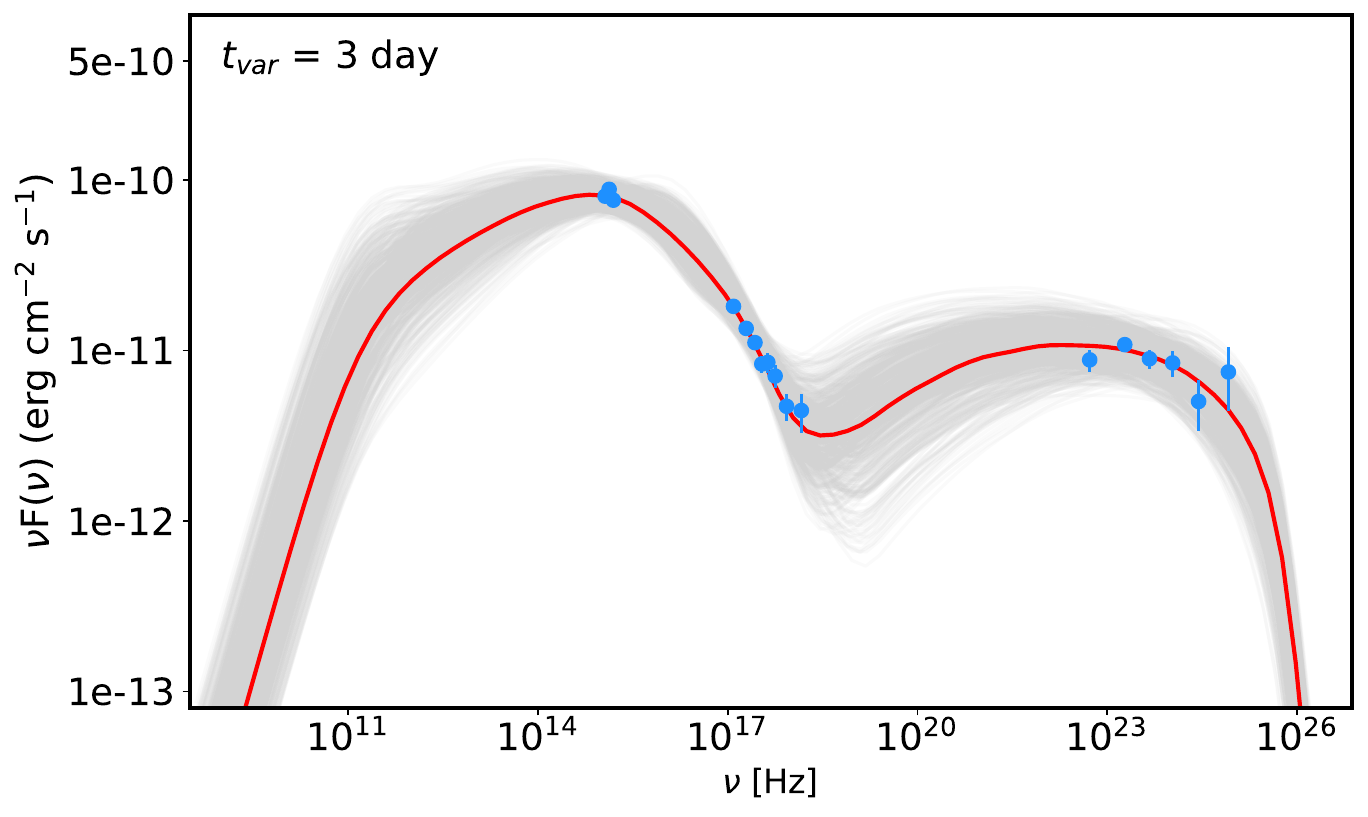}\\
    \includegraphics[width=0.48\linewidth]{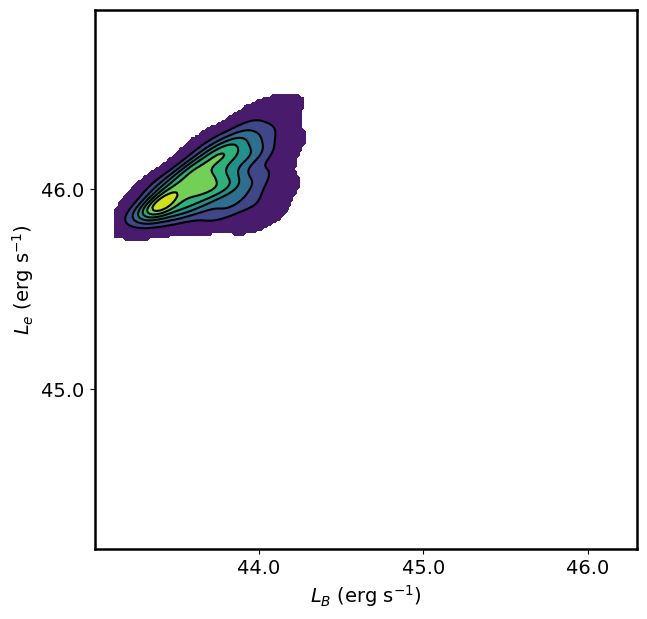}
    \includegraphics[width=0.48\linewidth]{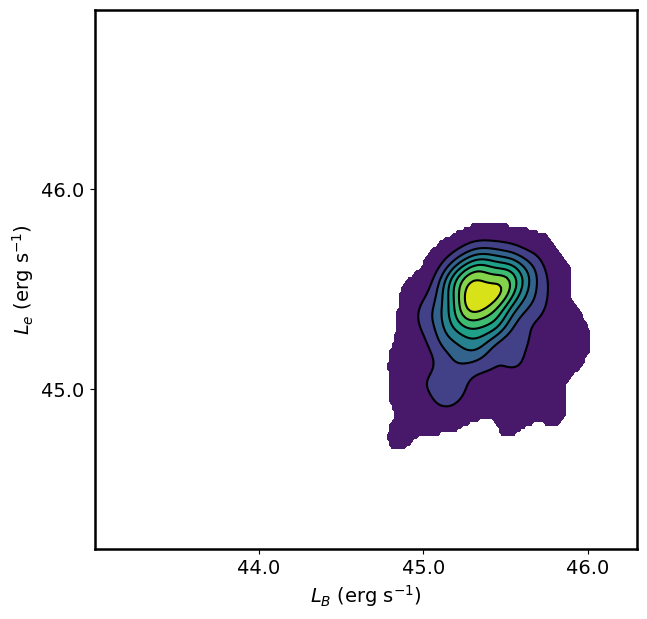}
    \caption{ Upper panel: The modeling of \source\ SED for the periods MJD 56748.60-56778.80 (left) and MJD 57490.73-57823.49 (right). The red line corresponds to the model at maximum likelihood, while the gray spectra represent the model uncertainty. The data points and their uncertainties are shown in blue. Extragalactic background light (EBL) absorption is taken into account in the model computation using the model of \citet{2011MNRAS.410.2556D}. Lower panels: Two-dimensional projection of the electron luminosity and magnetic luminosity corresponding to the SEDs shown in the upper panel. The scaling of these two plots is the same to show the different parameter space characteristic of these two SEDs.}
    \label{fig:cor}
\end{figure*}
 
\section{The origin of the broadband emission} \label{orig}

\subsection{Modeling method}

The double-peaked SED of BL Lac type blazars is typically interpreted within the framework of the SSC model
\citep{1985A&A...146..204G, 1992ApJ...397L...5M, 1996ApJ...461..657B}, to which we restrict our analysis. This
model assumes that the low-energy component is produced by synchrotron emission from electrons, while the HE
component is due to inverse Compton scattering of these photons on the same population of electrons (SSC).
Modeling the SEDs allows for the estimation of parameters describing the jet as well as
the emitting electrons, providing insights into the physical processes occurring in these sources
\citep[e.g.,][]{2008ApJ...686..181F,2020A&A...638A..14M, 2021MNRAS.502..836S, 2020MNRAS.496.5518S, 2011ApJ...736..131A}.
Contrary to previous studies, instead of selecting just one or a few SEDs for the modeling, almost all periods determined in
the temporal analysis and shown in the SED/LC animation that had sufficient data were modeled. The criteria for
selecting the SEDs were that data in both the low-energy and X-ray bands had to be available (since \gray\ data are available
by default) in order to shape both the low and HE components. In total, 736 SEDs over 802 met these criteria. We note that
hierarchical Bayesian method could allow to alleviate this criteria, but this works is outside the scope of the current paper.

The selected SEDs are modeled through \mmdc{} using the innovative method proposed in \citet{2024ApJ...963...71B}, which
employs a neural networks for modeling SEDs instead of a computationally expensive numerical model. This approach
enables self-consistent modeling of the broadband emission from blazars, taking into account both the injection and cooling
of particles. The details of the model and the methodology are provided in \citet{2024ApJ...963...71B} , while a similar
approach for the external inverse Compton model is described in \citet{2024ApJ...971...70S}. In the SSC
model, electrons are injected into a spherical region with a radius of $R$, which is filled with a homogeneous and constant
magnetic field $B$. The injection spectrum of the electrons is assumed to follow a power-law with an exponential cut-off,
\begin{align}
    \dot Q = Q_{e,0} \gamma^{-p} \exp \left( -\frac{\gamma}{\gamma_{\rm max}} \right)
\end{align}
when $\gamma > \gamma_{\rm min}$ and 0 otherwise. In this equation, $Q_{e,0}$ is the normalisation, $p$ is the electron power-law index, $\gamma_{\rm min}$ is the minimum electron Lorentz factor at injection (electrons can achieve a smaller Lorentz factor by cooling), and $\gamma_{\rm max}$ is the maximum electron Lorentz factor. The emission region is moving with relativistic velocities (Lorentz
factor $\Gamma$) toward the observer, who sees the jet at an angle $1/\Gamma$, resulting in the emission being Doppler
amplified by a factor of $\delta \equiv \Gamma$. Accordingly, the model has seven free parameters:
$p$, $\gamma_{\rm min}$, $\gamma_{\rm max}$, $B$, $R$, $\delta$, and the electron luminosity $L_{\rm e}$, such that

\begin{align}
    L_e = \pi R^2 \delta^2 m_e c^3 \int_1^{\infty} \gamma Q_e d\gamma
\end{align}
Considering a wide range for these parameters (see Table 1 in
\citet{2024ApJ...963...71B}), the temporal evolution of the electrons is followed, and the corresponding SEDs are computed
using soprano \citep{2022MNRAS.509.2102G}. A convolutional neural network (CNN)
is then trained on these data, producing the radiative signature of the electrons, which is used instead of the numerical model for modeling the SEDs.
   
\subsection{Modeling results} \label{res}

The selected broadband SEDs were fitted using MultiNest a nested sampling tool \citep{FHB09}, assuming 1500 active
points and a tolerance of $0.4$ to ensure efficient sampling and convergence. Additionally, as data in the energy
range $<10^{10}$ Hz are often unavailable, we set the minimum Lorentz factor $\gamma_{\rm min}=100$. To further
reduce the number of free parameters, we use the variability time to link the radius of the emitting region $R$ and the
Doppler boost $\delta$, such that $R=\delta\:c\:t_{\rm var}/(1+z)$. The variability timescale, $t_{\rm var}$,
is an unknown characteristics of the source that cannot be estimated for each considered SED.
Therefore, we fit the entire dataset using two values of $t_{\rm var}$, which are 1 day and 3 days.
In most cases, i.e. 559 out of 736 SEDs, $t_{\rm var}=3$ days provides satisfactory
results, but in a few cases (177 SEDs), the fit did not converge, so the results with $t_{\rm var}=1$ day were
used instead. The modeling of all 736 contemporaneous SEDs as an animation is available \href{https://youtu.be/oTGEZR4M3Ow}{on youtube} and the individual plots are available on \href{https://github.com/gevorgharutyunyan/OJ287/tree/a8279e3db7b62769689433ff9b08a1a57ea48ac0/sed}{github}. An animation showing the time evolution of the parameter posterior distributions is available on \href{https://youtu.be/VVe6bsKKmmY}{youtube} and the corresponding figures are available on \href{https://github.com/gevorgharutyunyan/OJ287/tree/a8279e3db7b62769689433ff9b08a1a57ea48ac0/posteriors}{github}.  

We show in Fig. \ref{fig:cor} the modeling results of two SEDs of \source\ for which it is in two different states. The model at maximum likelihood is represented in red, while the gray spectra
correspond to the model uncertainty. The SED presented on the left panel is best modeled with $p=2.12$,
$\gamma_{\rm max}=1.15\times10^4$, and $\delta=34.14$. The size of the emission region (connected to $\delta$,
assuming a 3-day variability) is $R=2.65\times10^{17}$ cm. The magnetic field is estimated to be
$B=1.12\times10^{-2}$G, resulting in the magnetic jet luminosity
\( L_{B}=\pi c R_b^2 \Gamma^2 U_{B} =\)$3.85\times10^{43}\:{\rm erg\:s^{-1}}$, which is lower than that of the
electron luminosity, $L_{\rm e}=2.12\times10^{46}\:{\rm erg\:s^{-1}}$.  Similarly, the parameters obtained from the
fit of the SED presented on the right panel of Fig. \ref{fig:cor} are $p=2.32$, $\gamma_{\rm max}=1.16\times10^5$,
$\delta=29.09$, and the emission is produced in a region with a radius of \( R = 2.26 \times 10^{17} \) cm. In this case,
the source was bright in the optical/UV band with the peak-flux of the low-energy component exceeding that of the HE component,
so the magnetic field is higher as compared with the previous case: $B=0.12$. The electron luminosity, in this
case \( L_{\rm e} = 1.10 \times 10^{45}\, {\rm erg\, s^{-1}} \) is a factor 2 lower than the magnetic
field luminosity \( L_{\rm B} = 2.18 \times 10^{45}\, {\rm erg\, s^{-1}} \). The SED modeling results presented in Fig.
\ref{fig:cor} indicate that, even when analyzing the SED of the same source, the radiative signature changes
significantly over time.

We note that the broadband SED of OJ 287 during various periods has been modeled using different datasets and/or different models with varying assumptions about the origin of the emission \citep[e.g.,][]{2018MNRAS.479.1672K, 2021A&A...654A..38P, 2024ApJ...973..134A}, among others. For example, a one-zone synchrotron/SSC model was used in \citet{2021A&A...654A..38P} to model \source\ emission by combining various X-ray states, while a multi-zone modeling approach was adopted in \citet{2024ApJ...973..134A} to account for the observed VHE \gray\ data. While a direct comparison of the obtained parameters is not possible due to differences in datasets and modeling assumptions, the estimated jet parameters (e.g., luminosities) are within the range inferred from the current study.

\subsection{Time-evolution of modeling parameters}

\begin{figure*}
    \centering
    \includegraphics[width=0.9\linewidth]{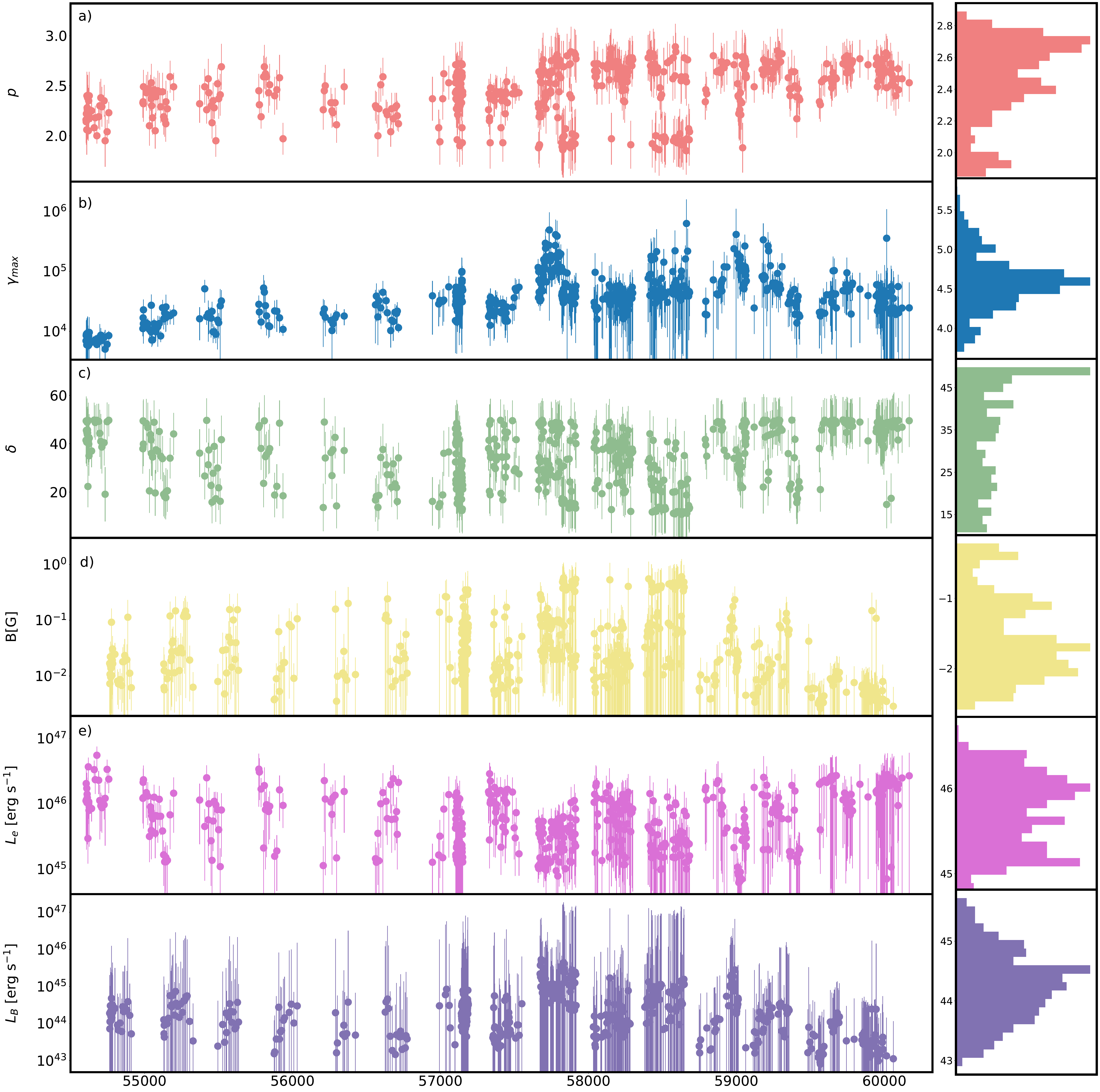}
    \caption{The temporal evolution of the model parameters derived from the modeling of the 559 "time-resolved" SEDs assuming $t_{\rm var} = 3$ days. The error bars indicate one-sigma uncertainty in the model parameter estimation.
    a) Power-law index of the emitting electrons. b) Cut-off energy of the emitting electrons. c) Doppler factor.
    d) Magnetic field strength in the emitting region. e) and f) Jet luminosity in the electron and magnetic field,
    respectively. The histograms in the right panel show the distributions
    of each parameter, enlightening the emission variability
    of \source{}, and how they translate to variation in the emission and jet parameters. }
    
    \label{fig:SED_params}
\end{figure*}

The modeling of the SEDs provides the model parameters for each period and facilitates the investigation of temporal
changes in the emission properties of \source. Fig. \ref{fig:SED_params} shows the temporal evolution
of $p$, $\gamma_{\rm max}$, $\delta$, $B$, $L_{\rm e}$, and $L_{\rm B}$ (for the evolution of model parameters derived from modeling 177 SEDs with \( t_{\rm var} = 1 \) day, see Fig. \ref{fig:SED_params_1}.). The high amplitude of parameter changes highlights the
variability in the physical processes contributing to the overall emission from \source. When considering only the parameters obtained with the assumption of a 3-day variability period, the power-law index of
the electron distribution varies within the range $p=1.85-2.89$ (Fig. \ref{fig:SED_params}, panel a), with a mean
value of $2.45$. The variation in $p$ could reflect changes in the efficiency of particle acceleration mechanisms at
work in the jet of \source. A decrease in $p$ might indicate a period of stronger acceleration, injecting more high-energy
particles into the emission region. There are 54 (29) periods out of 559 (177) modeled assuming $t_{\rm var}=3$ day (1 day) for which $p \leq 2.0$, whereas during other
periods, a softer spectrum is estimated, implying lower-energy particles dominate the electron energy distribution. Similarly, the cut-off energy $\gamma_{\rm max}$ varies from $5.05 \times 10^3$ to $6.27 \times 10^5$ ($4.24 \times 10^3$ to $2.26 \times 10^5$) for $t_{\rm var}=3$ days ($t_{\rm var}=1$ day), see Fig. \ref{fig:SED_params}, panel b. The evolution of $\gamma_{\rm max}$ provides insights into
the maximum energy that particles can achieve, higher estimates of $\gamma_{\rm max}$ might indicate stronger
acceleration mechanisms that enable particles to reach HEs, whereas lower estimates of $\gamma_{\rm max}$ could suggest that in
these periods energy losses due to synchrotron cooling or inverse Compton scattering dominate, preventing electrons from reaching HEs.

The modeling also provides information about changes within the jet. The observed fluctuations
of the magnetic field strength, $B$, which ranges from $2.61 \times 10^{-3}$ G to $0.604$ G ($9.42 \times 10^{-3}$ G to $0.997$ G) when $t_{\rm var}=3$ days ($t_{\rm var}=1$ day), see Fig. \ref{fig:SED_params}, panel d, may
indicate alterations in the magnetic field structure of the jet, possibly induced by turbulence that periodically injects energy
into the system. Such structural changes in the jet are also evident from variations in the Doppler boost $\delta$, see Fig. \ref{fig:SED_params}, panel
c, which ranges from 10.8 to 49.85 (13.05 to 49.98) for $t_{\rm var}=3$ days ($t_{\rm var}=1$ day). These changes may be explained by the shifts of the jet orientation or adjustments in the bulk
flow speed, potentially driven by instabilities within the jet. We note that the value of 49.98 is defined by the
range considered for $\delta$. In these cases, $\delta$ cannot be estimated accurately and instead reaches
the upper limit of the parameter space considered. Although a higher range for $\delta$ could be considered, such extreme
values are uncommon for blazar jets.

The time-evolution of the electron and magnetic luminosities,  $L_{\rm e}$ and $L_{\rm B}$, are shown in
Fig. \ref{fig:SED_params} panels e and f respectively, which reveal how power is
distributed between particles and the magnetic field, shedding light on the internal energy balance of the
jet. The electron luminosity $L_{\rm e}$ varies from $8.10 \times 10^{45} \, \mathrm{erg \, s^{-1}}$ to  $5.53 \times 10^{46} \, \mathrm{erg \, s^{-1}}$ (from $4.81 \times 10^{45} \, \mathrm{erg \, s^{-1}}$ to  $2.46 \times 10^{46} \, \mathrm{erg \, s^{-1}}$) for $t_{\rm var}=3$ days (1 day)
while $L_{\rm B}$ ranges from $2.29 \times 10^{44} \, \mathrm{erg \, s^{-1}}$ to $5.29 \times 10^{45} \, \mathrm{erg \, s^{-1}}$ (from $2.75 \times 10^{42} \, \mathrm{erg \, s^{-1}}$ to $3.36 \times 10^{45} \, \mathrm{erg \, s^{-1}}$) for $t_{\rm var}=3$ days (1 day).
The modeling indicates that, for $t_{\rm var}=3$ days ($t_{\rm var}=1$ day), there are 338 (119) periods such that the electron luminosity exceeds  the magnetic luminosity, $L_{\rm e} / L_{\rm B} > 10$, implying that the jet is matter-dominated
rather than Poynting-flux-dominated. This suggests that the kinetic energy of particles (electrons and possibly
protons) primarily drives the jet dynamics in the dissipation region. We note that in the current study, no assumptions were made regarding the proton content in the jet. The presence of protons is expected to significantly impact the jet energetics and, if they achieve a non-thermal distribution, could potentially contribute non-negligibly to the radiative output in the HE/VHE \gray\ bands \citep[e.g.,][]{2013ApJ...768...54B,2016ApJ...832...17B}.

In a smaller number of cases, namely 27 (11) for $t_{\rm var}=3$ days ($t_{\rm var}=1$ day), 
$L_{\rm e}/L_{\rm B} < 1$, indicating phases for which the magnetic luminosity temporarily dominates particle energy.
This magnetic dominance influences the flux of synchrotron emission by strongly cooling the electrons and contribute to variability, specifically in the optical bands. As an example, the lower panel of Fig. \ref{fig:cor} shows the two-dimensional posterior probability
distributions of $L_{\rm e}$ and $L_{\rm B}$  for the periods MJD 56748.6–56778.80, for which
$L_{\rm e}/L_{\rm B} \sim 550$, and MJD 57490.73–57823.49, for which $L_{\rm e}/L_{\rm B} \sim 0.5$.
In the first case, the peak flux of the synchrotron and inverse Compton
components are similar (left upper panel of Fig. \ref{fig:cor}), while in the second case, the synchrotron component
is significantly stronger (right upper panel of Fig. \ref{fig:cor}). An animation of the time- evolution  of the two-dimensional projections
of the posterior probability distributions of $L_{\rm e}$ and $L_{\rm B}$ is available on
\href{https://youtu.be/TYBAAwXftGM}{youtube} while the corresponding figures are available on \href{https://github.com/gevorgharutyunyan/OJ287/tree/a8279e3db7b62769689433ff9b08a1a57ea48ac0/correlations}{github}.

\section{Discussions and Conclusions} \label{disc}

In this paper, we presented a detailed analysis of the emission from \source{}, examining multi-wavelength
data accumulated from 2008 to 2023 and conducting in-depth modeling of its time-resolved multi-wavelength SEDs. The extensive
data analysis performed on \source{} across multiple wavelengths—optical/UV, X-ray, and \gray\ bands—is crucial for
building a comprehensive understanding of its emission characteristics and their evolution over time. The adaptively binned \gray\ light-curve reveals that the \gray\ flux was significantly enhanced during several periods.
Similarly, flares are also observed in the optical/UV band. Since the simultaneous multi-wavelength coverage of \source{} is not complete, specifically during flaring periods, we do not attempt to correlate intra-band flux variations, which will be the purpose of another study from another blazar. Yet, this  detailed multiwavelength analysis allowed us to construct and track the evolution of
the broadband emission from \source\ and to investigate temporal variations (in both flux and frequency) of the emission components. This forms a foundation for determining and interpreting the physical conditions driving the emission from \source.

\begin{figure}
    \centering
    \includegraphics[width=\linewidth]{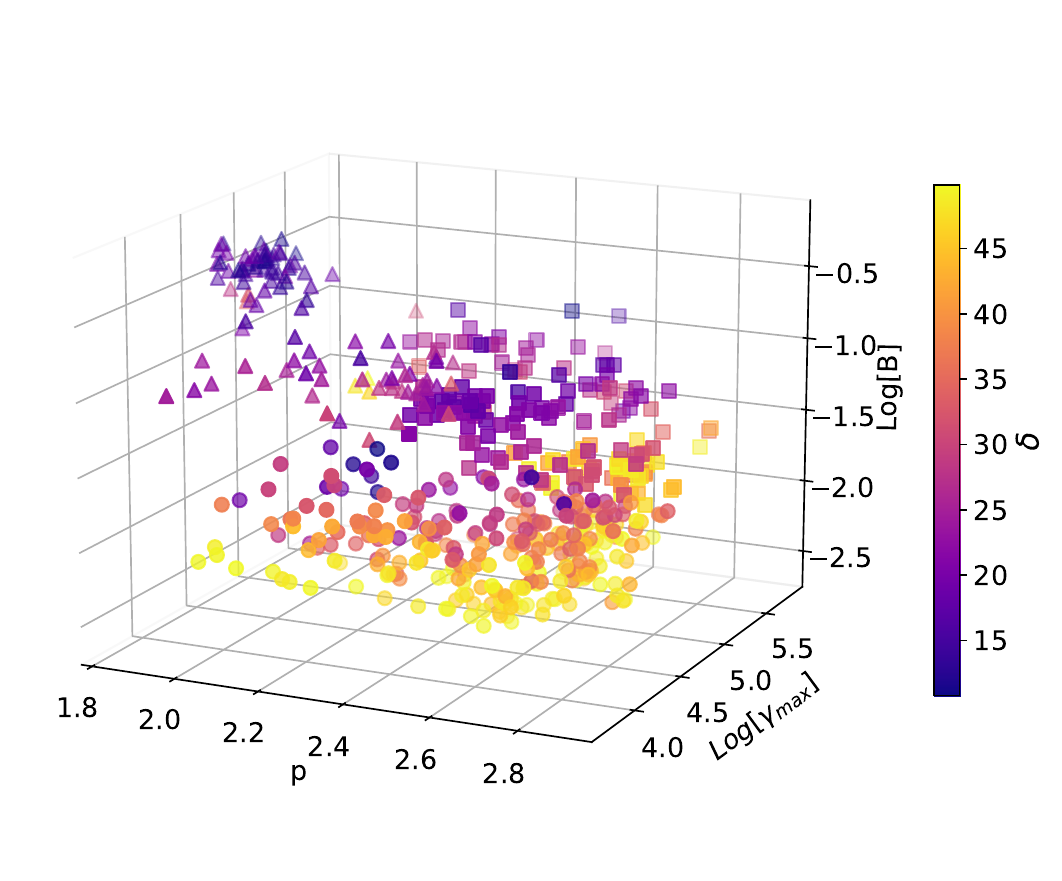}
    \caption{3D plot of the power-law index $p$ versus the magnetic field $B$ and the cut-off energy of the electrons $\gamma_{\rm max}$. The color coding indicates the Doppler factor. Only parameters for $t_{var} = 3$ days are displayed. A clustering analysis reveals the existence of three clusters in the retrieved parameters, represented as triangles, circles and squares.}
    \label{fig:3d_proj}
\end{figure}

The modeling of the 736 observed periods with contemporaneous data provides
the foundations for an in-depth study of the physical processes operating within the jet of
\source. The inferred variations in the model parameters highlight changes in the underlying
mechanisms contributing to particle acceleration and emission, as well as structural transformations within the jet.
For example, various acceleration mechanisms—such as shock acceleration under varying conditions, stochastic acceleration,
and magnetic reconnection—can account for the power-law index of the electrons, which varies within the range of 1.85–2.89.
Indices close to $\simeq 2.0 $ can be explained by non-relativistic shocks \citep[i.e., diffusive shock acceleration
][]{1978MNRAS.182..147B, 1978ApJ...221L..29B, 1987PhR...154....1B}, whereas softer spectra may form under weaker
shocks, relativistic shock conditions \citep{1987ApJ...315..425K, 1990ApJ...360..702E, 1998PhRvL..80.3911B, 2004APh....22..323E},
or magnetic reconnection \citep{2013MNRAS.431..355G, 2003ApJ...589..893L, 2005MNRAS.358..113L,2009MNRAS.395L..29G, 2016MNRAS.462.3325P}.

Restricting the discussion to spectra with assumed $t_{var} = 3$ days, we present in 
\ref{fig:3d_proj} a 3D plot of the power-law index of the emitting electrons as a function of the magnetic
field strength \(B\) and the cut-off energy $\gamma_{\rm max}$
of the electrons. The color coding indicates the Doppler factor
$\delta$. We applied k-min clustering algorithm using the scikit-learn package\footnote{https://pypi.org/project/scikit-learn/1.5.2/} to group data points based on similarity in the features. The algorithm finds three clusters, shown as squares, triangles and circles. The data points shown as circles are characterized by low values of $B < 0.02$G, soft  $p$ (with most values such that $p\gtrsim 2.4$), large $\gamma_{\rm max} > 3.16\times10^4$, and relatively high Doppler boosts ($>40$).
The squares represent data with intermediate $B \sim 0.03$G and $\gamma_{\rm max} \sim 3.16\times10^4$, soft $p \gtrsim 2.3$, and moderate Doppler boosting $<30$. These first two clusters display parameters usually obtained from blazar SED modeling.

In contrast to the first two clusters, the triangles have high $B >0.1$G, intermediate $\gamma_{\rm max} \sim 3.16\times10^4$, and harder $p \lesssim 2$ than the other two groups. The Doppler boosting for this last cluster falls in the intermediate range $\sim 30$.
The high magnetic field and  harder electron power-law indices for the cluster represented by triangles,
could indicate that under stronger magnetic confinement, electrons experience prolonged
interactions within acceleration sites, such as shocks, turbulence regions, or magnetic reconnection zones. This enhanced
confinement allows electrons to undergo multiple acceleration cycles. As they gain energy with each cycle, their energy
distribution hardens, resulting in a lower value for the power-law index $p$. However, an increase in magnetic
field strength also raises synchrotron cooling rates, $\propto B^2$, preventing the cut-off energy
\(\gamma_{\rm max}\) from reaching higher values, see e.g. \citet{dHM96}.
This balance between acceleration and cooling establishes a critical constraint within the jet. The interplay between magnetic
confinement, acceleration, and cooling effects provides insights into the variability of the emission spectrum over time. The observed correlation between high magnetic fields and hard electron spectra highlights the dynamic conditions within the jet, suggesting that localized magnetic enhancements could trigger transient episodes of intense acceleration, leading to observable changes in emission properties. Such episodes could result from magnetic reconnection events or the development of shock fronts within the jet flow. Additionally, the relatively low values of $\delta$ during these periods imply that the impact of jet orientation and bulk relativistic motion on the observed variability is minimal. The increase in emission intensity likely arises from local enhancements, in particle acceleration efficiency, rather than changes in the jet's orientation or speed.

One of the key outcomes of this study is the systematic analysis of the temporal evolution of
the broadband emission characteristics of \source\ using a large dataset acquired over a decade in the period 2008–2023.
This is unlike many previous investigations that relied on modeling a single-epoch or a time-averaged SEDs.
Classical single-snapshot SED modeling, while valuable, averages out the dynamical evolution of jet parameters and
physical processes, potentially missing key variability patterns. Instead, we modeled 736 epochs with our newly developed CNN-based model, trained on a kinetic model which accounts for electron
injection and cooling. We systematically determined the parameter space associated with diverse emission
properties, corresponding to diverse physical conditions in the jet. This approach enabled to track the time
evolution of these parameters, investigate temporal variations, and identify correlations, providing an unprecedented level of detail
in characterizing the jet activity of \source\ - insights that were previously inaccessible in static SED modeling. Indeed, this methodology reveals distinct clusters of parameters, clarifying how
different acceleration and cooling regimes manifest over time. In the case of \source, this allowed for the identification of emission
states characterized by unique combinations of magnetic field \( B \), electron index \( p \), and Doppler boost
\( \delta \), tied to underlying mechanisms such as varying acceleration processes (e.g., shocks, turbulence) and magnetic confinement.
Moreover, the detailed parameter sets derived in this work serve as valuable inputs for  numerical simulations of particle
acceleration. These datasets provide realistic initial conditions for modeling shock
formation, turbulence-driven acceleration, and magnetic reconnection processes in the jet. Thus, our methodology offers a robust framework
for interpreting multi-wavelength variability and paves the way for simulation-driven insights into the physical mechanisms governing the
jet activity of \source.

The modeling in this study represents a significant step forward, enabling nuanced, data-rich and comprehensive interpretations of blazar behavior over time. These results underscore the need for sustained multi-wavelength monitoring, as the diversity of observed states highlights the complex and rapidly evolving nature of emissions from blazars in general, and for OJ 287 in particular. This study not only enhances the understanding of \source\ but also lays a foundation for applying  detailed time-resolved modeling of blazar SEDs leading to a better understanding of their variable nature.

\section*{Acknowledgements}
We acknowledge the use of data, analysis tools and services from the Markarian Multiwavelength Data Center (\url{www.mmdc.am}), the Astrophysics Science Archive Research Center (HEASARC), the Fermi Science Tools, the Astrophysics Data System (ADS), and the National Extra-galactic Database (NED).\\
The research was supported by the Higher Education and Science Committee of MESCS RA (Research project No 23LCG-1C004).\\
This work used resources from the Armenian National Supercomputing Center (Aznavour supercomputer).

\section*{Data Availability}
All the data used in this paper is available from \mmdc{} (\url{www.mmdc.am}). The data is also available upon reasonable request to the corresponding author.



\bibliographystyle{mnras}
\bibliography{biblio} 




\appendix

\section{Evolution of modeling parameters when $t_{\rm var}=1$ day}
The evolution of the model parameters when SED modeling was performed with  $t_{\rm var} = 1$ day instead of  $t_{\rm var} = 3$ days.

\begin{figure*}
    \centering
    \includegraphics[width=0.9\linewidth]{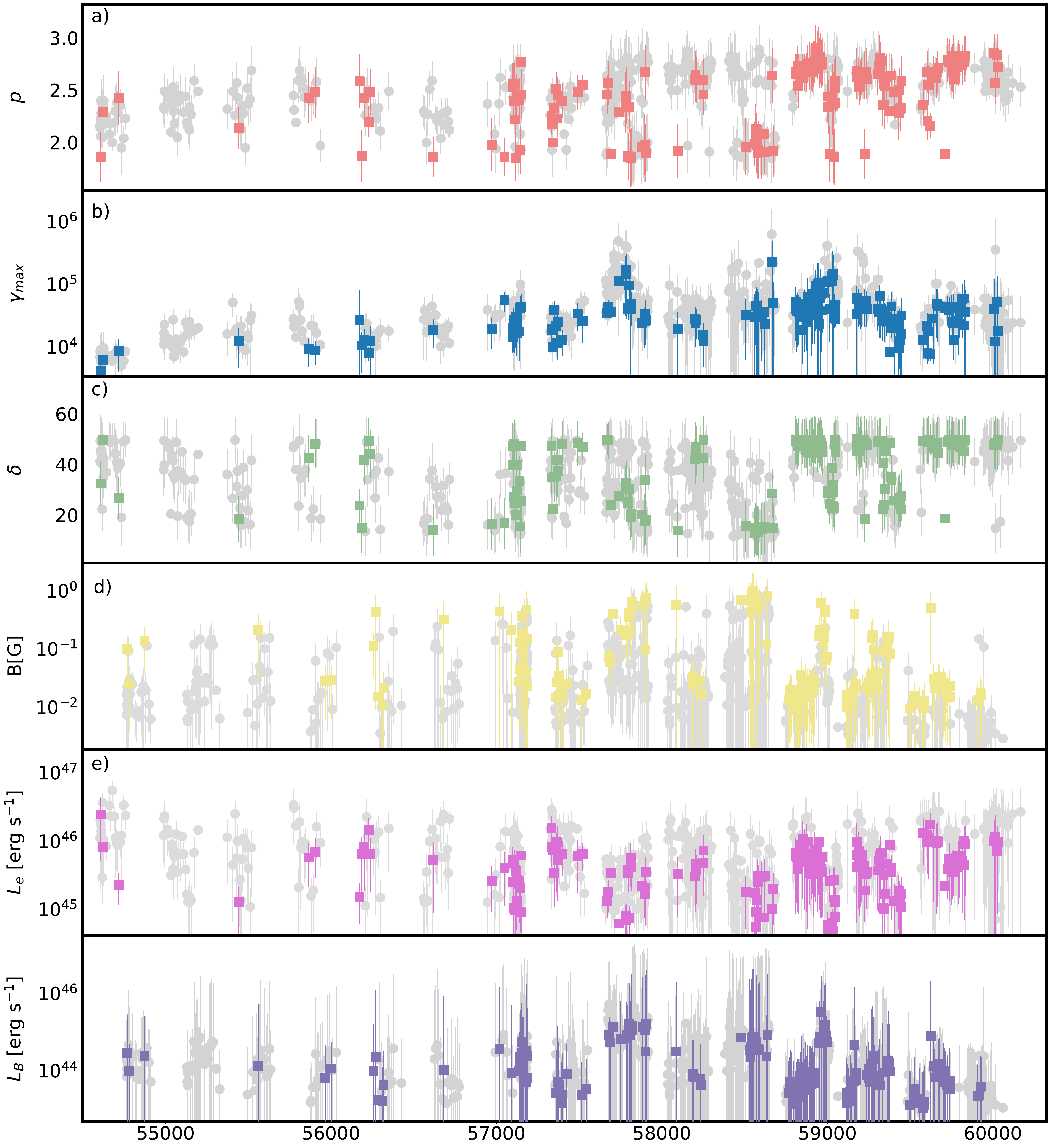}
    \caption{The temporal evolution of the model parameters derived for \( t_{\rm var} = 1 \) day. In the background, the results from Fig. \ref{fig:SED_params}, corresponding to \( t_{\rm var} = 3 \) days, are shown in gray for comparison. The parameters and color coding are the same as in Fig. \ref{fig:SED_params}}.
    
    \label{fig:SED_params_1}
\end{figure*}


\bsp	
\label{lastpage}
\end{document}